\documentclass[reprint,amsmath,amssymb,aps]{revtex4-1}

\usepackage{graphicx}
\usepackage{natbib}
\usepackage{amsmath}
\usepackage{gensymb}
\usepackage{enumerate}

\usepackage[usenames, dvipsnames]{color}

\newcommand\Beq{\begin{eqnarray}} 
\newcommand\Eeq{\end{eqnarray}}

\newcommand\Bfig{\begin{figure}} 
\newcommand\Efig{\end{figure}}

\renewcommand\L {\mathcal{L}}

\newcommand{\T}{{\cal T}}
\newcommand{\Z}{{\cal Z}}

\makeatletter
\let\Hy@backout\@gobble
\makeatother

\newcommand*{\GtrSim}{\smallrel\gtrsim}

\makeatletter
\newcommand*{\smallrel}[2][.8]{%
  \mathrel{\mathpalette{\smallrel@{#1}}{#2}}%
}
\newcommand*{\smallrel@}[3]{%
  \sbox0{$#2\vcenter{}$}%
  \dimen@=\ht0 %
  \raise\dimen@\hbox{%
    \scalebox{#1}{%
      \raise-\dimen@\hbox{$#2#3\m@th$}%
    }%
  }%
}
\makeatother

\begin{document}

\title{Connection between Nonlinear Energy Optimization and Instantons}

\author{Daniel Lecoanet}
\affiliation{%
Physics Department, University of California, Berkeley, CA 94720, USA\\
Astronomy Department and Theoretical Astrophysics Center, University of California, Berkeley, CA 94720, USA\\
Princeton Center for Theoretical Science, Princeton University, Princeton, NJ 08544, USA\\
Department of Astrophysical Sciences, Princeton University, Princeton, NJ 08544, USA\\
Kavli Institute for Theoretical Physics, University of California, Santa Barbara, CA 93106, USA}

\author{Rich R. Kerswell}
\affiliation{%
DAMTP, Centre for Mathematical Sciences, Cambridge University, Cambridge CB3 0WA, UK \\
School of Mathematics, University of Bristol, Bristol BS8 1TW, UK\\
Kavli Institute for Theoretical Physics, University of California, Santa Barbara, CA 93106, USA}
\date{\today}

\begin{abstract}

How systems transit between different stable states under external perturbation is an important practical issue. We discuss here  how a  recently-developed energy optimization method for identifying the minimal disturbance necessary to reach the basin boundary of a stable state is connected to the instanton trajectory from large deviation theory of noisy systems. In the context of the  one-dimensional Swift--Hohenberg equation which has multiple stable equilibria, we first show how the energy optimization method can be straightforwardly used to identify  minimal disturbances---{\em minimal seeds}---for transition to specific attractors from the ground state. Then, after generalising the technique to consider multiple, equally-spaced-in-time perturbations,  it is shown that  the instanton trajectory is indeed the solution  of the energy optimization method  in the limit of infinitely many perturbations provided a specific norm is used to measure the set of discrete perturbations.  Importantly, we find that the key features of the instanton can be  captured by a low number of discrete perturbations (typically one perturbation per basin of attraction crossed). This suggests a  promising new diagnostic for systems for which it may be impractical to calculate the instanton.

\end{abstract}

\maketitle

\section{Introduction}

How and when systems can transit between different stable states in the presence of ambient disturbances is of fundamental importance in  understanding their behaviour in practice. There are two clear limits which can be explored:  the system experiences just one finite-amplitude disturbance or is continuously perturbed by low amplitude noise. A technique for examining the  former scenario has recently been developed using a nonlinear energy optimization method \citep{pk10,cherubini10,mono11, Kerswell14, pringle15, Kerswell18} which identifies the disturbance of smallest  amplitude---the {\em minimal seed}---which can initiate the transition.   A promising application of this approach is to the problem of subcritical transition to turbulence in parallel shear flows where the minimal seeds which emerge are typically localized and therefore appear relevant to experimental studies \citep{mono11,pringle15}. In the latter, small-noise situation where the transition between different stable states is rare, large deviation theory is used to seek the most-likely transition  trajectory in the limit of zero noise known as the {\em instanton} \citep{fw98}.  One can use the instanton approach to identify the fast dynamics which lead to transitions over long timescales in fast-slow systems \citep[e.g.,][]{bouchet16,grafke16}.  Again, fluid dynamics has provided an important application area for these ideas with instantons computed in a number of different contexts \citep{bs09,bouchet11,grafke13,wan13, wan15}.  The purpose of this paper is to explore the connection between these two approaches by extending the nonlinear optimization method to treat multiple perturbations. The instanton approach should be a limiting case of the optimization method as the number of discrete perturbations becomes large under an appropriate norm. What is particularly interesting is to gain some insight into  how quickly this limit is approached as the number of discrete perturbations increases.

Rather than study the Navier--Stokes equations, we perform optimization calculations for the much simpler, one-dimensional Swift--Hohenberg equation (SH).  \citet{bk06} show that SH has multiple localized stable equilibria as a result of homoclinic snaking which provides a  richer phase-space environment in which to explore both approaches than the usual bistability of the Navier--Stokes equations  in, for example, shear flows \citep{pringle15, wan15}.  The existence of multiple attractors  opens up the possibility that optimal transition trajectories between any two states can  take non-trivial forms involving third-party  basins of attraction. SH has also been studied extensively \citep[][and references within]{Kao14}. 

This paper is organized as follows.  In section~\ref{sec:states} we describe the SH problem, the different equilibrium states present for our chosen parameters, and their properties.  Section~\ref{sec:minimum_seed} describes the minimal energy perturbations from the trivial state into any of the other stable states of the problem.  We are able to select for the different stable states by optimizing the time-averaged energy, because the stable states have sufficiently disparate energies.  Section~\ref{sec:instanton} extends the optimization calculations to include multiple perturbations and the calculation of the instanton.  The discretized instanton corresponds to the optimal set of perturbations which occur at every timestep of our simulation.  Finally we conclude in section~\ref{sec:conclusions}.

\section{Dynamics of the Swift--Hohenberg System}\label{sec:states}

We consider the one-dimensional Swift--Hohenberg equation (SH) with a quadratic--cubic nonlinearity,
\Beq\label{eqn:SHE}
\partial_t u + (1 + \partial_x^2)^2u - a u = 1.8 u^2 - u^3,
\Eeq
following \citet{Kao14}.  Different coefficients for the nonlinear terms---and different nonlinearities---will give similar properties \citep[e.g.,][]{bk06}.  The trivial state ($u=0$) is linearly stable for $a<0$ so we pick $a=-0.3$.  The primary instability of the system has wavenumber $k=1$, corresponding to a characteristic length of $L_c=2\pi$ so we consider a domain of length $6L_c$ to allow multiple equilibria.  All simulations are run using the open-source, pseudo-spectral code Dedalus\footnote{dedalus-project.org}\citep{burns18}.  The solutions are calculated as a Fourier expansion with 256 modes, and we use $\times 2$ padding to preventing aliasing errors on the grid from the cubic nonlinearity.  For timestepping, we treat the linear terms implicitly using backward Euler, and we treat the nonlinear terms explicitly using forward Euler, with a constant timestep of $0.1$ (the temporal resolution of the trajectories was verified by additional simulations with reduced timestep size).

%
%
\begin{figure}
  \centerline{\includegraphics[width=3.4in]{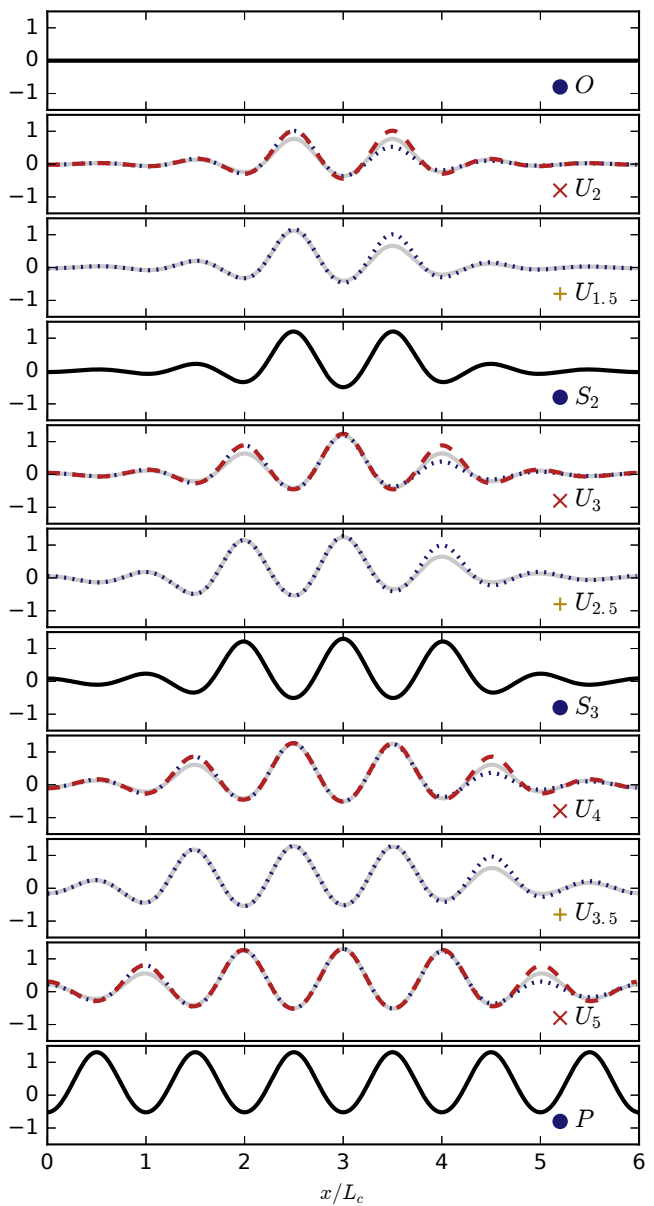}}
  \caption{The nonlinear solutions to SH (equation~(\ref{eqn:SHE})) with $a=-0.3$, shown in black (stable) or grey (unstable).  $U_{1.5}$, $U_{2.5}$, and $U_{3.5}$ are unstable edge states with only a single unstable eigenvector, and are not $\Z$-symmetric.  $U_2$, $U_3$, $U_4$ and $U_5$ are $\Z$-symmetric unstable solutions with two unstable eigenvectors.  We plot the sum of the solution and $\Z$-symmetric eigenvectors (with some small amplitude) in red dashed lines, and the sum of the solution and eigenvectors without $\Z$ symmetry in blue dotted lines.}
\label{fig:solutions}
\end{figure}

Our choice of $a=-0.3$ has four stable solutions, and several unstable solutions.  The solutions are shown in figure~\ref{fig:solutions}.  The energy of each solution is given in table~\ref{tab:energies}.  The four stable solutions are the trivial state at the origin, $O$, the periodic state $P$, and two localized states, $S_2$ and $S_3$, which have two and three large amplitude maxima ($u\GtrSim 1$).  Although all the states are periodic with length $6L_c$, we call $P$ the periodic state because it also has periodicity of $L_c$.  This choice of parameters has enough different states for the optimization problem to give non-trivial results, but not so many states to obfuscate the analysis.

The equations have reflection and translation symmetries of which two,
\Beq
\Z:&\quad& x \rightarrow 6L_c - x, \\
\T:&\quad& x\rightarrow x+\frac{L_c}{2} \mod \, \,6L_c,
\Eeq
are important for the discussion which follows although none of our calculations are restricted to any symmetric subspace.
Because we have defined our solutions as centered around $x=3L_c$, the stable states as well as $U_2$, $U_3$, $U_4$, and $U_5$ are $\Z$-symmetric.  These unstable states have one $\Z$-symmetric unstable eigenvector, and one $\Z$-antisymmetric unstable eigenvector.  The other unstable states, $U_{1.5}$, $U_{2.5}$, and $U_{3.5}$ lack $\Z$ symmetry.

%
%
\begin{figure*}
  \centerline{\includegraphics[width=7.1in]{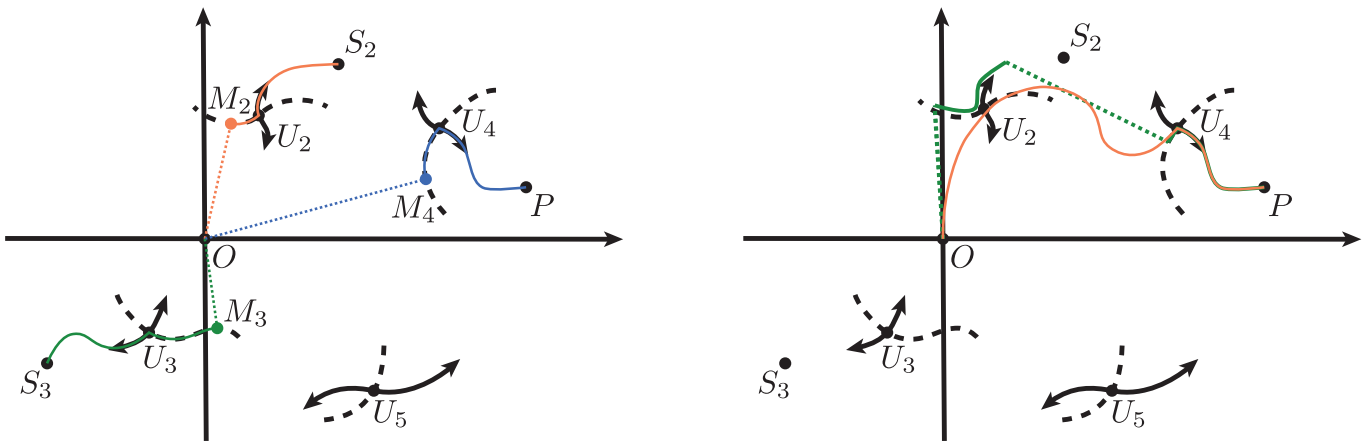}}
  \caption{Schematic diagram of the $\Z$-symmetric manifold.  The dashed lines show basin boundaries, and the unstable states are drawn with their unstable manifold.  The unstable states are edge states of the $\Z$-symmetric dynamics. Left: The trajectories of the minimal seeds for each stable state are shown in different colors.  The dotted line corresponds to a perturbation, and the solid colored line corresponds to the evolution of SH.  The minimal seed is the closest point on the basin boundary to $O$. Right: The trajectory of the optimal set of two perturbations (green), and the instanton (orange).  Because we fix the time between the two perturbations, the first perturbation for the green curve does not go to $M_2$.}
\label{fig:schematic}
\end{figure*}

Figure~\ref{fig:schematic} shows a schematic depiction of the $\Z$-symmetric manifold.  Although $U_2$, $U_3$, $U_4$, and $U_5$ have two unstable eigenvectors for the full problem, they only have a single unstable eigenvector in the $\Z$-symmetric subspace, and thus are edge states.  $U_2$ separates $O$ from $S_2$; $U_3$ separates $O$ from $S_3$; $U_4$ separates $S_2$ from $P$; and $U_5$ separates $S_3$ from $P$.  Although we perform our optimization in the full phase space (i.e. no symmetries are imposed on the dynamics), we find that the optimal perturbations satisfy $\Z$ symmetry, so the their dynamics lie on the $\Z$-symmetric manifold.

\begin{figure*}
  \centerline{\includegraphics[width=7.1in]{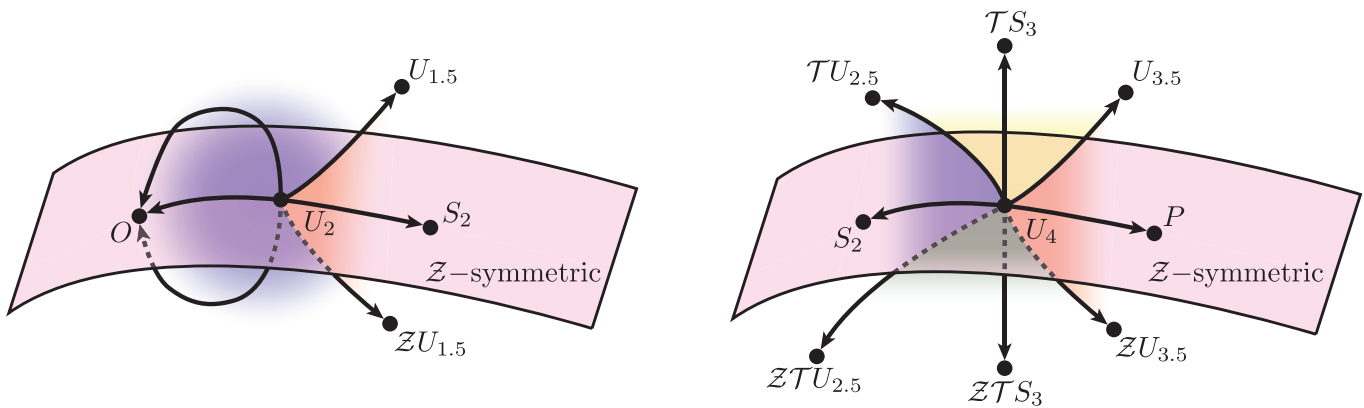}}
  \caption{Schematic diagram of the phase space around $U_2$ (left) and $U_4$ (right).  Both states have two unstable eigenvectors---one tangential to the $\Z$-symmetric manifold, and one directed out of  the manifold.  The arrows are all in the two-dimensional unstable manifold of $U_2$ and $U_4$, and the  colors correspond to the basin of attraction of the different stable states within the unstable manifold.
}
\label{fig:U2_U4}
\end{figure*}

In full phase space,  $U_{1.5}$ is an edge state between $O$ and $S_2$; $\T U_{2.5}$ ($U_{2.5}$ shifted by $L_c/2$ in $x$) is an edge state between $S_2$ and $\T S_3$; and $U_{3.5}$ is an edge state between $\T S_3$ and $P$.  The two dimensional unstable manifolds of $U_2$ and $U_4$ are  depicted in figure~\ref{fig:U2_U4} (those for $U_5$ and $U_3$ mimick $U_2$ and $U_4$ respectively).  $U_2$ has an unstable $\Z$-asymmetric eigenvector (blue dotted line in figure~\ref{fig:solutions}) which leads back to $O$.  A linear combination of the two unstable eigenvectors leads to the edge states $U_{1.5}$ and $\Z U_{1.5}$.  The $\Z$-asymmetric unstable eigenvector of $U_4$ leads to either $\T S_3$ or $\Z \T S_3$.  Because the unstable manifold contains four stable states, it also contains four saddle states---$\T U_{2.5}$, $\Z \T U_2.5$, $ \Z U_{3.5}$ and $U_{3.5}$---each positioned between a given neighbouring pair of stable states.

%
%

In the remainder of the paper, we quantitatively compare the states and different trajectories.  To aid in this comparison, the state $u$ is projected onto two coordinates: the total energy per characteristic length, and the energy in the third through fifth Fourier mode per characteristic length,
\Beq
E_t(u) &=& \frac{1}{6}\int \frac{1}{2}|u|^2 \ dx \nonumber \\
& = & \frac{1}{6}\left(\frac{1}{2}\hat{u}(0)^2 + \sum_{k=1}^{127} |\hat{u}(k)|^2\right), \\
E_{3-5}(u) & = & \frac{1}{6} \left(|\hat{u}(3)|^2 + |\hat{u}(4)|^2 + |\hat{u}(5)|^2\right),
\Eeq
where $\hat{u}$ denotes the spatial Fourier transform of $u$, and the $k\neq 0$ Fourier modes are multiplied by two due to Hermitian symmetry.  Other  choices of coordinates give similar plots but  $E_t$ and $E_{3-5}$ seemed the best at separating the different states in the plane.

The partitioning of phase space into the various basins of attraction  is key to understanding the minimum energy perturbations that lead to each of the different stable solutions to SH.  In the next section, we will find that these states are on the stable manifold of the unstable solutions $U_i$.

\section{Minimal Seed Perturbations}\label{sec:minimum_seed}

We now carry out nonlinear optimization calculations to calculate the minimal seed for the stable states $S_2$, $S_3$ and $P$.  The minimal seed is the minimum energy perturbation from $O$ which evolves into each of these stable states.  We will refer to the minimal seeds as $M_{2}$, $M_{3}$ and $M_P$.  This is a first step in considering multiple perturbations as well as continuous perturbations (section~\ref{sec:instanton}).

To find the minimal seeds, we calculate the perturbation with fixed energy $E_0$ which maximizes the time-integrated energy
\Beq\label{eqn:objective}
F[u(t)] = \int_{0}^{t_f} \, \int_{0}^{6L_c} \ \frac{1}{2}  |u|^2 \, dx dt.
\Eeq
We do this with an iterative approach (derived in appendix~\ref{sec:appendix}):
\begin{enumerate}
\item Integrate $u$ from $t=0$ to $t=t_f$, including the perturbation $\delta u$ at $t=0$;
\item Initialize the adjoint variable $\beta(x,t_f)=0$ at $t=t_f$;
\item Integrate the adjoint variable according to the adjoint equation
\Beq\label{eqn:adjoint}
\partial_t\beta - (1+\partial_x^2)^2\beta + a \beta = -3.6u\beta + 3 u^2\beta + u
\Eeq
{\em back} to $t=0$;
\item Update the perturbation $\delta u$ according to
\Beq
\delta u(x) \rightarrow \delta u(x) + \epsilon [\,\alpha \delta u(x) - \beta(x,0)\, ],
\Eeq
where $\epsilon=0.073$ is a small parameter setting the size of the update and $\alpha$ is a Lagrange multiplier used to enforce the constraint that the perturbation has initial energy $E_0$.
\end{enumerate}
\noindent The adjoint equation is evolved in time using Dedalus, with the same numerical choices as the integration of SH.  This algorithm can be repeated until we find a local maximum of the time-integrated energy.

%
%
\begin{figure*}
  \centerline{\includegraphics[width=7.1in]{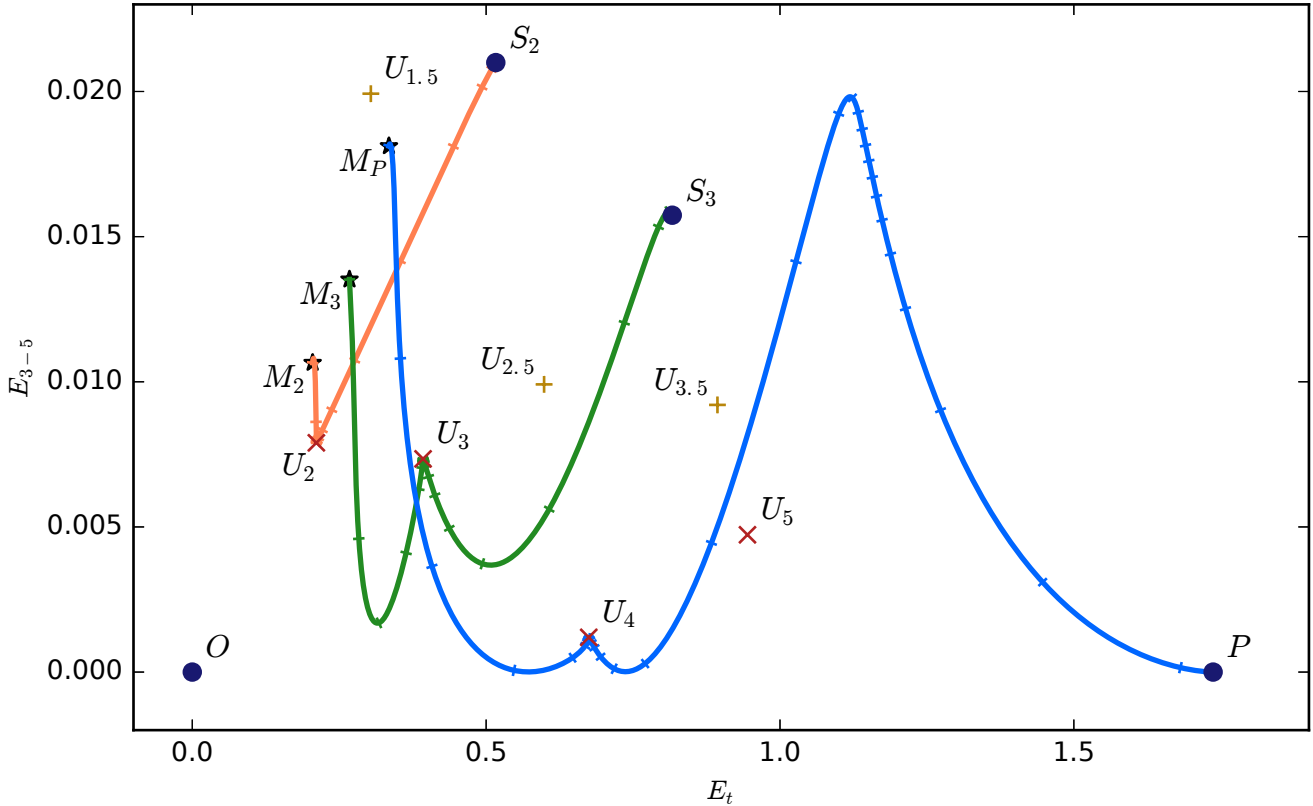}}
  \caption{Minimum energy perturbations to $O$ to the three other stable states, $S_2$, $S_3$, and $P$ (shown with black stars).  The trajectories are also plotted, showing that in each case, the minimal seed is on the stable manifold of one of the $\Z$-symmetric unstable states, $U_2$, $U_3$, or $U_4$.  Tick marks are placed on the trajectories every 5 time units.  See figure~\ref{fig:minimum_seeds} for each of the perturbations.}
\label{fig:phase_space_pert}
\end{figure*}

The algorithm depends on many choices.  We use a final time $t_f=50$, which is long enough to reach the stable states $O$, $S_2$, and $S_3$, or to get close to the solution $P$.  Using a later final time would lead to better estimates for the minimum seeds but also makes the optimization procedure more sensitive to the perturbations and hinders convergence \citep{Kerswell14}.
The use of the time-integrated energy (see~(\ref{eqn:objective})) rather than the more usual final energy as our objective function is motivated by  optimization calculations involving  multiple perturbations (described in the next section). With multiple perturbations, maximizing the time-integrated energy rather than the energy at the final time $t_f$ encourages the algorithm to introduce large perturbations at $t=0$, rather than wait some amount of time before perturbing the system (which is equivalent to optimizing over fewer perturbations). Some calculations were nevertheless done with the final energy as the objective function and  found to produce similar minimum seeds albeit with slower convergence.

Trajectories which approach a given stable solution have larger time-integrated energies than trajectories which approach lower energy solutions allowing minimal seeds for each to emerge naturally as $E_0$ is increased.  To do this, the optimization procedure is started with white noise of energy  $E_0=E_i$ much greater than the energy of the minimal seed.  Then the optimization loop is run for $E_0'<E_0$  for up to two hundred iterations to see if the system is still in the attractor of the desired state.  If it is, the optimal perturbation is rescaled down in energy again  and the optimization loop repeated.  If the system is not in the attractor of the desired state, the energy of the optimal perturbation is either rescaled upwards $E_0'>E_0$, or the optimization is restarted with white noise of the same energy.  Using this procedure, we calculate the energy of the minimal seed to within an energy per characteristic length ($E_t$) tolerance of $5\times 10^{-4}$.

The procedure is repeated  hundreds of times until we have several perturbations with the same low energy which are in the attractor of the desired state.  For state $S_2$, most initial noise guesses converge to the same low energy, whereas for state $S_3$, we converged to the lowest energy perturbation only 18 times after over 600 initial guesses.  Each of these perturbations are slightly different, as their energy is slightly larger than the energy of the minimal seed (given our tolerance of $5\times 10^{-4}$).  To get a better estimate of the minimal seed, we rescaled the perturbations to slightly lower amplitudes to see the minimum energy necessary to reach the desired state.

Although our optimization calculations do not impose $\Z$ symmetry, in each case, we find the perturbations are very close to being symmetric.  If the perturbation is symmetrized, we find that we can reach the desired state with slightly lower energies than by using the rescaled outputs of the optimization calculation.  Thus, we believe the minimal seeds are $\Z$-symmetric states.

Each of our target states $S_2$, $S_3$, and $P$ are well-separated in energy, so it is straightforward to calculate minimal seeds for each state individually by changing the energy of the initial perturbation.  Because of this, we were able to use the same objective function (see~(\ref{eqn:objective})) to find all three target states.  In other problems where different target states have similar energies, it may be more efficient to find the minimal seeds by varying the objective function.

%
%
\begin{figure}
  \centerline{\includegraphics[width=3.4in]{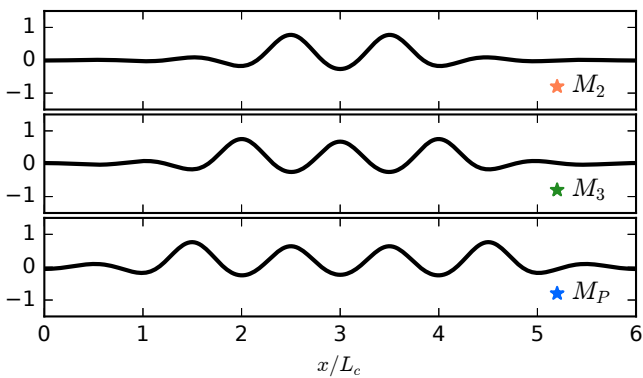}}
  \caption{The minimal seeds leading to stable states $S_2$, $S_3$, and $P$.  They each evolve toward one of the $\Z$-symmetric unstable states ($U_2$, $U_3$, or $U_4$) before reaching the desired stable state.}
\label{fig:minimum_seeds}
\end{figure}

The minimal seeds and the trajectories to their respective stable solutions are shown in figures~\ref{fig:phase_space_pert} \& \ref{fig:minimum_seeds}.  The total energy of each minimal seed is given in table~\ref{tab:energies}.  The minimal seeds and their trajectories lay on the $\Z$-symmetric manifold, and the trajectories are depicted heuristically in the left panel of figure~\ref{fig:schematic}.  The minimal seed is the closest point of approach between $O$ and the stable manifold of the unstable states $U_2$, $U_3$, and $U_4$, which are each edge states of the $\Z$-symmetric problem.  It is worth remarking that $U_5$ is also an edge state of the $\Z$-symmetric problem, but has higher energy than $U_4$, so one would expect its stable manifold to be further from $O$ than $U_4$'s stable manifold (although this does not have to be true).

\section{Multiple Perturbations and Instantons}\label{sec:instanton}

In the previous section, we found the optimal single perturbation to state $O$ which led to another stable state.  We now consider $n$ perturbations $\delta u_1$, $\delta u_2$, $\ldots$, $\delta u_n$ which act at times $t_1=0$, $t_2$, $\ldots$, $t_n$.  This is a discretized version of the continuous forcing problem,
\Beq\label{eqn:forced}
\partial_tu + (1+\partial_x^2)^2u  - au - 1.8u^2 +u^3 = f(x,t).
\Eeq
In the limit of large $n$, with perturbations which are equally spaced in time by $\Delta t$, we can approximate $f(x,t_i)\approx \delta u_i/\Delta t$. If the system is forced with low amplitude white noise, i.e., $f(x,t)=\sqrt{\epsilon}dW(x,t)$, where $dW$ is a Wiener process in time and space, then the probability to transition between states is
\Beq
p\sim \exp(-I[u]/\epsilon),
\Eeq
where the action 
\Beq
I[u] =\int_0^T \, \int_0^{6L_c} \frac{1}{2} |f|^2 \, dx dt
\Eeq
\citep{fw98}. The instanton trajectory, $u_I(x,t)$, is the trajectory which starts and ends at the chosen stable states and corresponds to a noise sequence which minimizes the action (i.e. is most likely).  See appendix~\ref{sec:app-instanton} for more details about instantons.

When optimizing over multiple perturbations, we use a norm which will converge to the action $I$ in the limit of infinitely many perturbations,
\Beq\label{eqn:norm}
N\left[\left\{\delta u_i\right\}_{i=1}^n\right] = n\sum_{i=1}^n E_t(\delta u_i).
\Eeq
For a single perturbation, this is simply the energy of that perturbation (the norm used in the previous section).  In the limit of infinitely perturbations which are equally spaced in time, we have
\Beq
I[f] &=& \int_{0}^{t_f} \, \int_0^{6L_c} \frac{1}{2} |f(x,t)|^2 \, dx dt \nonumber \\
 &\approx& \sum_{i=1}^n \Delta t \int_0^{6L_c} \frac{1}{2} |f(x,t_i)|^2 \, dx = \sum_{i=1}^n \int_0^{6L_c} \frac{|\delta u_i |^2}{\Delta t} \, dx \nonumber \\
&=& \frac{6}{t_f} n \sum_{i=1}^n E_t(\delta u_i) = \frac{6}{t_f}N\left[\left\{\delta u_i\right\}_{i=1}^n\right], \label{eqn:instanton_norm}
\Eeq
where $n=t_f/\Delta t$, and the approximation becomes an equality in the limit $\Delta t\rightarrow 0$.  Thus, the minimal seed ($n=1$) and instanton ($n=\infty$) can be viewed as two extremes of the general optimization problem for arbitrary $n$. It may seem like a more natural choice of norm would have been the sum of the energies of the perturbations ($N(\{\delta u_i\})/n$) but this goes to zero as $n \rightarrow \infty$ (see table~\ref{tab:energies}) rather than tending to the finite limit like the chosen norm (\ref{eqn:norm}).

In this section, we calculate the optimal set of two, five, and five hundred perturbations.  The optimal set of five hundred perturbations corresponds to adding a perturbation at every time step and  so is the discretized instanton.  We call the perturbations associated with the instanton $\delta u_{I}$, and the optimal set of $n$ perturbations $\delta u_{nP}$.  We only calculate these for the transition between $O$ and $P$.  For simplicity, the perturbations are assumed to be equally spaced in time, with  $t_i=t_f (i-1)/n$, so the two perturbations in $\delta u_{2P}$ act at $t=0$ and $25$, and the five perturbations in $\delta u_{5P}$ act at $t=0$, $10$, $20$, $30$, and $40$.

The calculation is based on a generalization of the optimization algorithm described in section~\ref{sec:minimum_seed} (see appendix~\ref{sec:appendix}).  We optimize over a set of $n$ perturbations $\{\delta u_i\}$ with fixed norm $N_0$ to maximize the objective function given in equation~(\ref{eqn:objective}).  The only differences are that  steps 1.~and 4.~are replaced by
{ \renewcommand\labelenumi{\theenumi.$'$}
\begin{enumerate}
\item Integrate $u$ from $t=0$ to $t=t_f$, including the perturbations $\delta u_i$ at $t=t_i$;
\end{enumerate}}
\noindent and
{ \renewcommand\labelenumi{\theenumi.$'$}
\begin{enumerate}\addtocounter{enumi}{3}
\item Update the set of perturbations $\delta u_i$ according to
\Beq\label{eqn:update_multi}
\delta u_i (x)\rightarrow \delta u_i (x)+ \epsilon (n \alpha\delta u_i (x) - \beta(x,t_i) ),
\Eeq
where $\epsilon=0.073$ (or $0.018$ for the instanton calculation) is a small parameter setting the size of the update, and $\alpha$ is the single Lagrange multiplier used to enforce that the set of perturbations has norm $N_0$.
\end{enumerate}}

As for the single perturbation problem, we initialize the algorithm with random noise for all perturbations. Then the optimization procedure is repeated up to two hundred times to try to find a set of perturbations with norm $N_0$ that leads to $P$.  We then vary $N_0$ to find $\delta u_I$ ($\delta u_{nP}$), up to norm of $0.025$ ($5n\times 10^{-4}$).  We repeat this for about one thousand random initial conditions.  This gives several slightly different optimals which have the same norm (up to the tolerance).  To determine the best, we uniformly rescale the set of perturbations to slightly lower amplitudes, and see which set of perturbations can transition to $P$ at the lowest amplitude.  We also symmetrize $\delta u_I$ and $\delta u_{5P}$ ($\delta u_{2P}$ was already symmetric) to give our best estimate for the optimal set of perturbations. It's worth remarking that this strategy for finding the instanton is not the usual direct one of minimizing the action across all trajectories which connect $O$ and $P$.  Instead, the action is fixed and then the time-integrated energy of the system maximised to find a trajectory connecting $O$ and $P$.  The action is then systematically reduced until no such connection can be found anymore. The success of this  indirect approach  relies on the fact that the optimization algorithm will find a connection if possible at a given action, as this maximizes the time-integrated energy.
The equivalence of the approach used here and the usual instanton calculation is discussed in  appendix~\ref{correspondence} where a formal connection between the two variational problems is made.

%
%
\begin{figure*}
  \centerline{\includegraphics[width=7.1in]{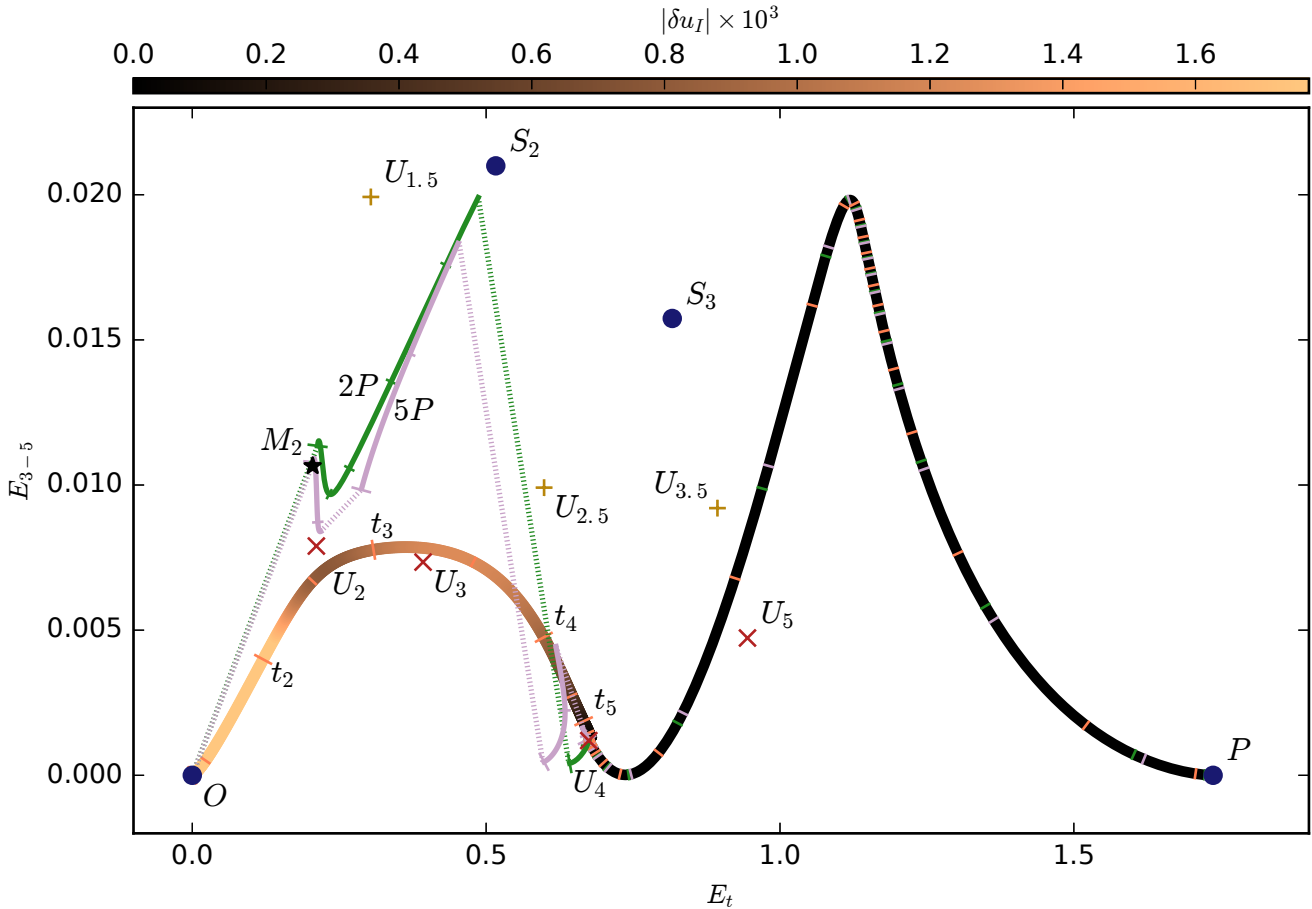}}
  \caption{Instanton trajectory (yellow to black line), and trajectories for optimal set of two and five perturbations to transition from $O$ to $P$.  The instanton trajectory's color corresponds to the size of the perturbation $\delta u_I(t)$ at that position in the trajectory.  The perturbations $\delta u_{2P,i}$ and $\delta u_{5P,i}$ are shown in dotted lines.  Tick marks are placed on each trajectory every five time units.  Long tick marks denote states and perturbations which are plotted in figure~\ref{fig:instantons}.  After they reach $U_4$, all three trajectories are identical, so they are all denoted with the black line. The trajectories associated with the optimal set of two and five perturbations are very close to each other, but are different from the instanton trajectory, or the minimal seed trajectory $M_P$ (figure~\ref{fig:phase_space_pert}).}
\label{fig:phase_space_instanton}
\end{figure*}

%
%
\begin{figure}
  \centerline{\includegraphics[width=3.4in]{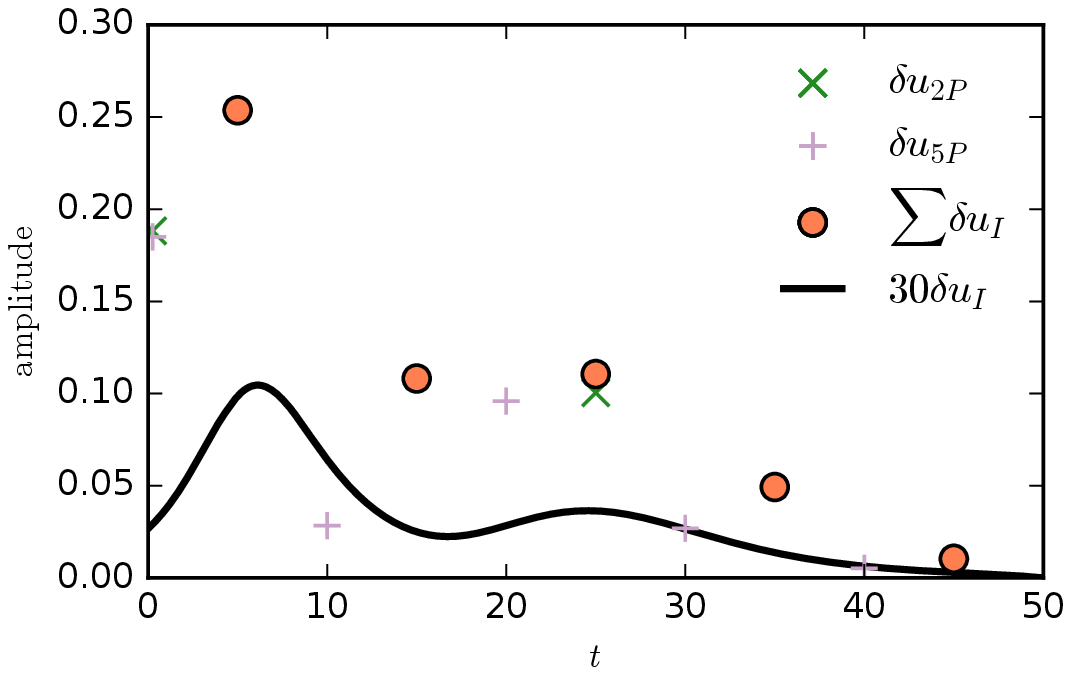}}
  \caption{The amplitude of the instanton's perturbation $\delta u_I(t)$ (black line), and each of the optimal perturbations $\delta u_{2P}$ and $\delta u_{5P}$ at the times of the perturbation.  Also shown is the amplitude of the sum of $\delta u_I$ between $t=10(i-1)$ and $10i$ (orange circles).
The largest perturbations in all cases are near $t=0$ and near $t=25$.  This corresponds to perturbing the system toward $U_2$, and then subsequently perturbing the system toward $U_4$.}
\label{fig:amplitudes}
\end{figure}

The right panel of figure~\ref{fig:schematic} shows a schematic depiction of the optimal set of two perturbations and the instanton.  The optimal set of two perturbations consists of a perturbation toward the stable manifold of $U_2$, followed by a second perturbation to the stable manifold of $U_4$, which leads to $P$.  The instanton trajectory approaches $U_2$, flows toward $S_2$, and then moves toward $U_4$.  In this sense, one can think of the instanton as primarily consisting of two ``types'' of perturbations, similar to the optimal set of two perturbations.  This is because the basin of attraction of $S_2$ separates the basins of attraction of $O$ and $P$.  Thus, our results suggest that one might expect the number of perturbations required to approximate the instanton may match the number of basins of attraction which need to be crossed.

\begin{figure}
  \centerline{\includegraphics[width=3.4in]{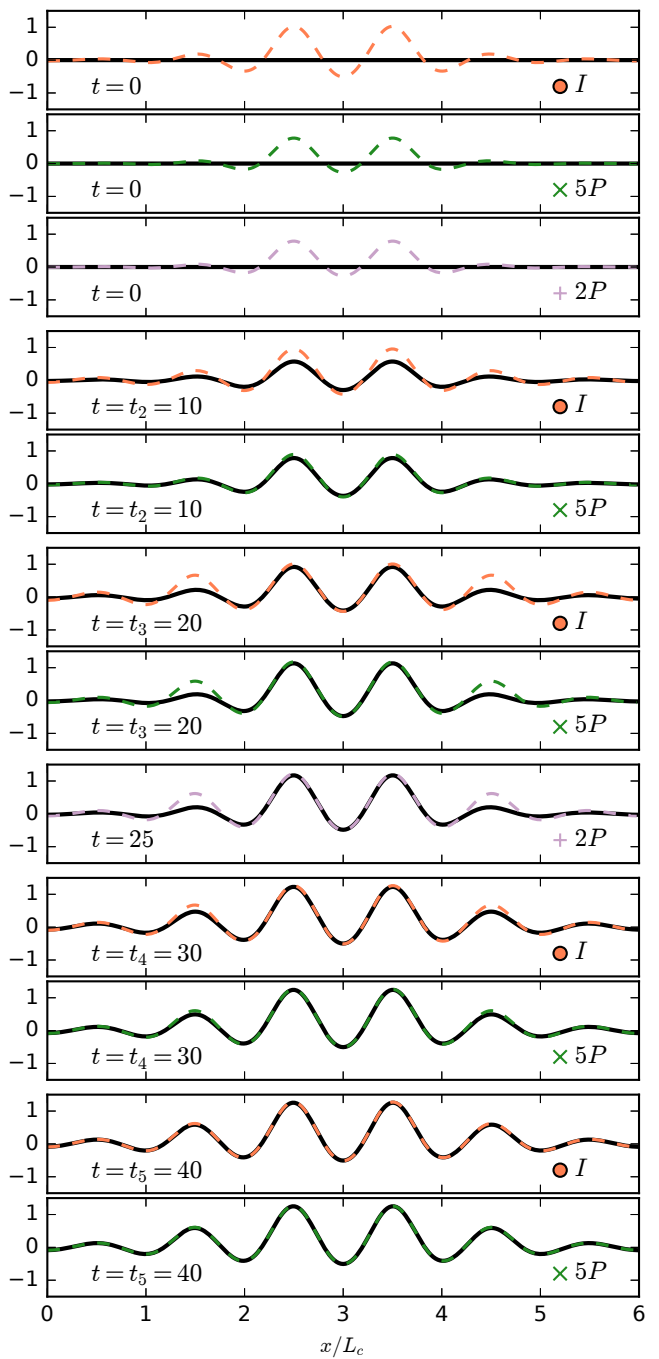}}
  \caption{The solution along the instanton trajectory ($I$) and along the trajectory associated with the optimal set of two and five perturbations ($2P$ and $5P$).  $I$ is shown every $10$ time units in black, with the dashed orange line showing the solution plus the sum of the perturbations over the next $10$ time units.  For $2P$ and $5P$, we show the solution right before each perturbation (in black), as well as right after each perturbation (in green or pink; dashed).  In all cases, initially the system develops two central large amplitude maxima, followed by two outer medium amplitude maxima.}
\label{fig:instantons}
\end{figure}

More quantitatively, figure~\ref{fig:phase_space_instanton} shows the instanton and the trajectories associated with $\delta u_{2P}$ and $\delta u_{5P}$, in the same projection as figure~\ref{fig:phase_space_pert}.  We will refer to the instanton trajectory as $I$, and the trajectory associated with $\delta u_{2P}$ and $\delta u_{5P}$ as $2P$ and $5P$.  We plot the solution and perturbations at different times in figure~\ref{fig:instantons}.  The color of the instanton trajectory in figure~\ref{fig:phase_space_instanton} corresponds to the size of the perturbation $\delta u_I(t)$ at each point on the trajectory (so the required noise is initially large to escape $O$'s basin of attraction and then vanishes once the system is in $P$'s the basin of attraction).  We measure the amplitude of the perturbation using
\Beq
|\delta u| = \sqrt{\frac{E_t(\delta u)}{6}},
\Eeq
the square root of the energy per unit length.  We use the amplitude (rather than the energy) because the amplitude of the sum of many perturbations in the same direction is equal to the sum of the amplitudes.  The amplitude of the perturbation as a function of time is shown in figure~\ref{fig:amplitudes}.

Initially, the instanton moves away from $O$ due to large amplitude perturbations producing two medium amplitude maxima in the center of the domain (see figure~\ref{fig:instantons}).  This lasts until $t\sim 15$, when the solution approaches the unstable $\Z$-symmetric state $U_2$.  The largest amplitude perturbations occur at early times because the system starts at a strong attractor ($O$).  Between $t\sim15$ and $t\sim 30$, the perturbation amplitude increases again, to perturb the system toward $U_4$.  Now the perturbations are predominately on two outer maxima, while the two central maxima grow in amplitude due to the flow of the system.  After $t=30$, the solution approaches $U_4$ without needing significant perturbations.  The sum of the perturbations from $t=40$ to $50$ shown in figure~\ref{fig:instantons} is so small it barely be seen by eye.  Although the instanton appears to pass close to $U_3$ (figure~\ref{fig:phase_space_instanton}), this is an artifact of our projection, as the solution is always negative at the center of the domain (at $x=3L_c$).

The trajectories $2P$ and $5P$ are similar to each other, as well as to the instanton.  In both cases, there are only two large perturbations, one toward $U_2$, and one to $U_4$.  Because there is only one basin of attraction between $O$ and $P$, having more than two perturbations does not change the result of the optimization significantly.

%
%

In section~\ref{sec:minimum_seed}, we found the minimal seed for $P$ has much lower energy than $U_4$.  However, the optimal set of multiple perturbations never approaches this minimal seed because the distance between $S_2$ and $U_4$ is smaller than the distance between $S_2$ and $M_P$.  This is because $U_4$ has two large amplitude central maxima, just like $S_2$, whereas $M_P$ has only medium sized central maxima.  By perturbing toward $M_2$, flowing toward $S_2$, and then perturbing close to $U_4$, the optimal set of multiple perturbations can take advantage of the energy-enhancing flow toward $S_2$.

Although the instanton follows a similar heuristic strategy as the optimal set of multiple perturbations, its trajectory using our projection is different from $2P$ and $5P$.  This is because the instanton perturbations enhance the outer two amplitude maxima at early times (see $t=0$ and $t=10$ in figure~\ref{fig:instantons}).  This moves energy from the fourth to second Fourier mode, decreasing $E_{3-5}$ relative to $2P$ and $5P$.

The instanton can enhance the two outer amplitude maxima at early times because the amplitude of its perturbations is larger than the amplitude of $\delta u_{2P}$ or $\delta u_{5P}$.  Figure~\ref{fig:amplitudes} shows that the sum of the amplitude $\delta u_I$ over time intervals of $10$ time units (orange circles) is always larger than the amplitudes of $\delta u_{2P}$ or $\delta u_{5P}$ at similar times.

The norm of the optimal set of perturbations increases as the number of perturbations increases.  If this trend occurs in other problems, it suggests that optimizing over a finite set of perturbations may give a lower bound on the norm of the instanton.  This should simplify calculations as optimizing over fewer perturbations is generally easier than calculating the instanton which has many more  degrees of freedom.

\begin{table}[b]
\caption{Energy of each of the solutions and minimal seeds.  For each optimal set of perturbations, we report the sum of the energy of the perturbations, as well as the norm (equation~(\ref{eqn:norm})). \label{tab:energies}}
\begin{tabular}{c|c|c}
state or perturbation  & $\sum E_t$ & $N=n\sum E_t$  \\[3pt] \hline
$O$ & 0 & \\
$S_2$ & 0.5164 & \\
$S_3$ & 0.8167 &  \\
$P$ & 1.737 &   \\
$U_{1.5}$ & 0.3038 &  \\
$U_{2.5}$ & 0.5986 &  \\
$U_{3.5}$ & 0.8936 & \\
$U_2$ & 0.2111 & \\
$U_3$ & 0.3927 & \\
$U_4$ & 0.6746 & \\ 
$U_5$ & 0.9447 &  \\ \hline
$M_2$ & 0.2048 & \\
$M_3$ & 0.2675 & \\
$M_P$ & 0.3346 & \\ \hline
$\delta u_{2P}$ & 0.2733 & 0.5465 \\
$\delta u_{5P}$ & 0.2700 & 1.350 \\
$\delta u_I$ & 0.0060 & 2.977 \\
\end{tabular}
\end{table}

\section{Conclusions}\label{sec:conclusions}

We have presented a series of optimization calculations using  the one-dimensional Swift--Hohenberg equation (SH) with a quadratic-cubic nonlinearity. Parameters such as  the domain length were chosen so that there are four stable solutions: the trivial solution $O$, two localized solutions with two or three large amplitude maxima ($S_2$ and $S_3$), and a global state $P$ which is periodic on the characteristic lengthscale.  There are also several symmetric and non-symmetric unstable solutions which are on the boundary between basin of attraction of the different stable solutions.

First we calculated the minimal seeds for transition from $O$ to either $S_2$, $S_3$, or $P$.  These are the smallest energy perturbation which causes transition  to the appropriate stable solution.  Geometrically, the minimal seed is the point of closest approach to $O$ on the basin boundary of each stable solution (left panel of figure~\ref{fig:schematic}).  In each case, the minimal seed is on the stable manifold of one of the symmetric unstable states (figure~\ref{fig:phase_space_pert}).  It is straightforward to find the minimal seeds for the various stable solutions because they are well separated in energy which forms the basis of the objective functional used. 

Next, we calculated the optimal set of multiple perturbations which guide the system from $O$ to $P$.  Mathematically, this is a straightforward modification to the optimization algorithm, but in practice the optimization problem is now more difficult because there are more perturbations to consider.  Using a special norm, we then calculated the optimal set of two perturbations ($\delta u_{2P}$), the optimal set of five perturbations ($\delta u_{5P}$), and the instanton ($\delta u_I$) in which the perturbations are a continuous function of time (i.e., optimizing over perturbations at every timestep).  
The trajectories for these three calculations are shown in figure~\ref{fig:phase_space_instanton}.  In all cases, we found that the easiest way to transition from $O$ to $P$ is to:  1.~Introduce two medium amplitude maxima in the center of the domain; 2.~Let the flow of SH grow these into two large amplitude maxima; 3.~Perturb the system to add two outer medium amplitude maxima (toward the unstable solution with four medium and large amplitude maxima, $U_4$); and 4.~Let the flow of SH lead to $P$. Importantly, even the two-perturbation optimal  captured the key features of the more involved instanton trajectory.

By generalising the recently-developed energy optimization technique to multiple perturbations {\em and} identifying the appropriate norm to measure a sequence of discrete perturbations, we have  established a formal link to the instanton trajectory of large deviation theory  which gives the most likely transition path between two stable states in noisy systems. What has emerged in doing this is the possibility that an optimization calculation incorporating only a very small number of discrete perturbations can give significant insight into the instanton trajectory. For the SH problem treated here, we found that  just two perturbations were enough to give a trajectory similar to the instanton because only two basins of attraction needed to be crossed (the basin of attraction of $S_2$ is between the basin of attractions of $O$ and $P$).  Clearly, more complicated problems with additional intervening basins of attraction will require more perturbations to approximate the instanton but this will be clear by gradually increasing the number of allowed perturbations in the optimization procedure (e.g. here $\delta u_{5P}$ is very similar to $\delta u_{2P}$).  

An optimal set of multiple perturbations  should also be  a good starting point for the calculation of an instanton and thereby lead to faster convergence than, say, random perturbations as an initial guess. Furthermore, it seems that the norm (equation~(\ref{eqn:norm})) of the optimal set of multiple perturbations gives a lower bound to the action of the instanton.  If this is true more generally, it may provide an interesting upper bound on the transition probabilities of systems under low amplitude noise without the need to calculate the full instanton.

\section*{Acknowledgments}
We thank Cedric Beaume for insight into properties of the Swift--Hohenberg equation, as well as Neil Balmforth and Stefan Llewellyn-Smith for helpful discussions.  DL is supported by a Hertz Foundation Fellowship, the National Science Foundation Graduate Research Fellowship under Grant No. DGE 1106400, a PCTS fellowship, and a Lyman Spitzer Jr.~fellowship. This work was initiated as a Woods Hole Oceanographic Institute Geophysical Fluid Dynamics summer project. Part of this work was completed at the Kavli Institute of Theoretical Physics program on Recurrent Flows: The Clockwork Behind Turbulence (Grant No. NSF PHY11-25915).

\appendix

%
%
\section{Derivation of the Optimization Algorithm}\label{sec:appendix}

We want to maximize the objective function $F[u(t)]$ defined in equation~(\ref{eqn:objective}) subject to the following constraints.  We require $u$ to satisfy SH, with perturbations $\delta u_i$ acting at times $t_i$, for $i=1,\ldots,n$.  We also require that ${\delta u_i}$ satisfy a norm condition $N[\delta u_i]=N_0$ (equation~(\ref{eqn:norm})).  To impose these constrains, we split $u(t)$ into $n$ different functions, $u_i(t)$, each of which are defined on $t\in [t_i,t_{i+1}]$.  For simplicity of notation, we also define $u_0=0$ and $t_{n+1}=t_f$.  Then we can define a Lagrangian
\Beq
\mathcal{L} &=& F[u(t)] + \alpha \left(N[\delta u_i]-N_0\right) \nonumber \\
 &+& \sum_{i=1}^{n}\int_0^{6L_c} dx \, \gamma_i(x) \left[u_{i}(t_i)-u_{i-1}(t_i)-\delta u_i\right] \nonumber \\
&+& \sum_{i=1}^{n} \int_{t_i}^{t_{i+1}} dt \int_0^{6L_c} dx \, \beta_i(x,t) \nonumber \\  
&\times& \left[\partial_t u_i + (1+\partial_x^2)^2 u_i - a u_i - 1.8 u_i^2 + u_i^3\right],
\label{A1}
\Eeq
where $\alpha$, $\gamma_i(x)$, and $\beta_i(x,t)$ are Lagrange multipliers imposing our constraints.

To maximize $\mathcal{L}$, we must vary the Lagrangian with respect to each of the variables.  Varying $\alpha$ imposes the norm condition, varying $\gamma_i$ imposes the perturbations, and varying $\beta_i$ requires $u_i$ to satisfy SH.  Varying with respect to $u_i$ gives the adjoint equation
\Beq
\partial_t \beta_i - (1+\partial_x^2)^2\beta_i &+& a \beta_i = \nonumber \\
&& - 3.6 u_i \beta_i + 3 u_i^2 \beta_i + u_i, \label{eqn:adjoint equation}
\Eeq
where the last term comes from our objective function.  Now we need a relation to relate the different $\beta_i$ to each other.  Varying with respect to $u_n(t_f)$ gives $\beta_n(t_f) = 0$.  Varying with respect to $u_i(t_i)$ gives $\gamma_i - \beta_i(t_i) = 0$, and varying with respect to $u_{i-1}(t_i)$ gives $-\gamma_i + \beta_{i-1}(t_i) = 0$, assuming $i > 1$.  Thus, we have that $\beta_{i-1}(t_i) = \beta_{i}(t_i)$; that is, $\beta$ can be viewed as a continuous variable satisfying the adjoint equation from $t=t_f$ to $t=0$.

Finally, we update the perturbations $\delta u_i$ in the direction
\Beq
\frac{\partial \mathcal{L}}{\partial \delta u_i} = \alpha n \delta u_i - \gamma_i = \alpha n \delta u_i - \beta(t_i).
\Eeq

%
%
\section{The Instanton}\label{sec:app-instanton}

An instanton is a trajectory which starts and ends at two chosen states which minimizes the action
\Beq
I[u] = \int_0^T \, \int_0^{6L_c} \frac{1}{2} |f|^2 \, dx dt,
\Eeq
where $f(x,t)$ is the forcing function (see equation~(\ref{eqn:forced})\,).  Here we are interested in transitions between $O$ and $P$.  Associated with the action is a Lagrangian,
\Beq
&&\mathcal{L}_I[u,\partial_tu] = \\
&& \int_0^{6 L_c} \frac{1}{2} \left|\partial_t u + (1+\partial_x^2)^2 u - a u - 1.8 u^2 + u^3\right|^2 dx. \nonumber
\Eeq

The conjugate momentum is
\Beq
p = \frac{\partial\mathcal{L}_I}{\partial \dot{u}} = \partial_t u + (1+\partial_x^2)^2 u - au - 1.8u^2 + u^3,
\Eeq
i.e., the forcing function $f$ (where $\dot{u}=\partial_t u$).  Then the \textit{instanton Hamiltonian} is
\Beq
&&\mathcal{H}_I[u,p] := \, \int^{6L_c}_0 \,p \dot{u}\, dx\,- \mathcal{L}_I \,=\\
&&\int_0^{6L_c} \left\{\frac{1}{2}p^2 - p\left[(1+\partial_x^2)^2u-au-1.8u^2+u^3\right]\right\} dx. \nonumber
\Eeq
The associated Euler-Lagrange equations are
\Beq
\partial_tu = p - \left[(1+\partial_x^2)^2u-au-1.8u^2+u^3\right], \\
\partial_tp = (1+\partial_x^2)^2 p - ap - 3.6pu + 3pu^2.
\Eeq
The first equation is the evolution equation for the system (equation~\ref{eqn:forced}).  The second equation is the unforced adjoint equation (equation~\ref{eqn:adjoint equation}). For more information about instantons and large deviation theory, we direct interested readers to \citet{laurie15}, and references within.

%
%

\section{Correspondence between $u_I$ and the Instanton}\label{correspondence}

The multiple perturbation approach is to find
\begin{equation}
\min_{N_0} \max_{\delta u_i} \, \mathcal{L}(\delta u_i, N_0)
\label{C1}
\end{equation}
where  $\mathcal{L}$ is defined in (\ref{A1}) and the outer minimization is performed over all $N_0$ which possess trajectories connecting the states $O$  and $P$.  The role of the objective functional $F$ is to ensure that such trajectories are found if they exist at a given $N_0$, but its precise form becomes increasingly unimportant as the minimum of $N_0$ is approached since the set of competitor trajectories shrinks down to one. The easiest way to see this mathematically is to rescale  and rewrite $\mathcal{L}$ as follows
\Beq
\mathcal{L}^* &:=&\mathcal{L}/\alpha = N[\delta u_i]+\frac{1}{\alpha} \left( F[u(t)]- \alpha N_0\right) \nonumber \\
 &+& \sum_{i=1}^{n}\int_0^{6L_c} dx \, \frac{\gamma_i(x)}{\alpha} \left[u_{i}(t_i)-u_{i-1}(t_i)-\delta u_i\right] \nonumber \\
&+& \sum_{i=1}^{n} \int_{t_i}^{t_{i+1}} dt \int_0^{6L_c} dx \, \frac{\beta_i(x,t)}{\alpha} \nonumber \\  
&\times& \left[\partial_t u_i + (1+\partial_x^2)^2 u_i - a u_i - 1.8 u_i^2 + u_i^3\right]. \label{C2}
\Eeq
The objective functional is then $N[\delta u_i]$ subject to the constraint that $F[u(t)]=\alpha N_0$ along with the other constraints. Minimizing this over $\delta u_i$ with the requirement that trajectories link the states $O$ and $P$ is the instanton calculation, albeit with this extra constraint. If the sensitivity of the minimum to this constraint is to vanish then $\alpha \rightarrow \infty$.  Empirically, we find that $\alpha$ increases as we approach the instanton. It is also clear here that $\beta_i$ must scale with $\alpha$ as the optimum is approached. This means that the homogeneous solution for $\beta$ in (\ref{eqn:adjoint equation}) increasingly dominates over the particular integral forced by the $F$-dependent inhomogeneous term (here $u_i$) so that
\begin{equation}
n \delta u_i \, \rightarrow \,f_i=p_i \, \leftarrow \, \frac{\beta_i}{\alpha} \quad \& \quad \alpha \rightarrow \infty
\end{equation}
as the optimum is approached. This establishes the correspondence.

\begin{figure}
  \centerline{\includegraphics[width=3.4in]{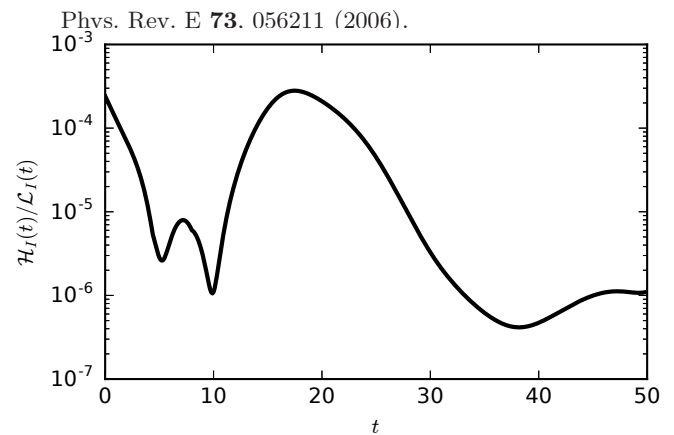}}
  \caption{The instanton Hamiltonian (normalized by the instanton Lagrangian) as a function of time.  The perturbations only act until $t_f=50$, so $\mathcal{H}_I$ is identically zero at later times.  The typical size of the terms in the Hamiltonian are given by $\mathcal{L}_I$, but they largely cancel out.  Thus, the Hamiltonian is very nearly constant, showing that the associated trajectory is an instanton.}
\label{fig:hamiltonian}
\end{figure}

An independent check is to show that  the Hamiltonian $\mathcal{H}_I$ of the instanton trajectory calculated using the optimization procedure is constant over time. This constant should be zero as once the system reaches the attractor of $P$, there is zero forcing, i.e., $p=0$, so $\mathcal{H}_I=0$ then.  In figure~\ref{fig:hamiltonian} $\mathcal{H}_I(t)$ is plotted normalised by $\mathcal{L}_I(t)$ which shows that $\mathcal{H}_I(t)$ is indeed small and so our trajectory $I$ approximates the instanton.

\bibliography{she}

\end{document}